\documentclass[twocolumn,preprintnumbers,amsmath,amssymb]{revtex4}
\usepackage{graphicx}
\usepackage{amsmath}
\usepackage{amssymb}
\usepackage{amsthm}
\usepackage{amsfonts}
\usepackage{enumerate}
\usepackage{overpic}
\usepackage{centernot}
\usepackage{tikz}

\usepackage{ulem}

\begin{document}

\title{Dissipationless topological quantum computation for Majorana objects in sparse-dense mixed encoding  process}

 \author{Ye-Min Zhan$^{1,2}$}
 \author{Guan-Dong Mao$^{1,2}$}
 \author{Yu-Ge Chen$^3$}
 \author{ Yue Yu$^{1,2}$}
 \thanks{Correspondence to: yuyue@fudan.edu.cn}
 \author{Xi Luo$^4$}
\thanks{Correspondence to: xiluo@usst.edu.cn}

 \affiliation {1. State Key Laboratory of Surface Physics, Fudan University, Shanghai 200433, China\\
 	2. Department of Physics,  Fudan University, Shanghai 200433, China\\
	3. Institute of Physics, Chinese Academy of Sciences, Beijing 100190, China\\
	4. College of Science, University of Shanghai for Science and Technology, Shanghai 200093, China }
\date{\today}

\begin{abstract}  
Topological quantum computation based on Majorana objects is subject to a significant challenge because at least some of the two-qubit quantum gates rely on the fermion (either charge or spin) parity of the qubits. This dependency renders the quantum operations involving these gates probabilistic when attempting to advance quantum processes within the quantum circuit model. Such an approach leads to  {significant} information loss whenever measurements yield the undesired fermion parity. To resolve the problem of  wasting information, we devise topological operations that allow for the non-dissipative correction of information from undesired fermion parity to the desired one.   {We  {will use} the sparse-dense mixed encoding process for the controlled-NOT gate as an example to explain how corrections can be implemented without affecting the quantum information carried by the computational qubits. This correction process can be applied  {to} either the undesired input qubits or the fermion parity-dependent quantum gates, and  {it} works for both Majorana-zero-mode-based and Majorana-edge-mode-based topological quantum computation.} 
 \end{abstract}

\maketitle

\section{ Introductions}

 {The discovery of the topological state of matter has shed new light on designing the next generation of low-energy-consuming quantum electronic and spintronic devices \cite{topo1,topo2,topo6,tosp1,tosp2,topo3,topo4,topo5}. Among these devices, the potential applications of topological quantum computation based on Majorana objects have attracted  a lot of research interest. }
The study of Majorana physics in condensed matter physics stems from Moore-Read's wave function for the $\nu=5/2$ even denominator fractional quantum Hall effect \cite{MR}, or  generally the $p$-wave superfluid/superconductor \cite{RG2000}. These Majorana objects obey the non-Abelian statistics, which can be equivalently described by Majorana zero modes (MZMs) \cite{NW, Inv}.   Following Kitaev's anyon-based fault-tolerant quantum computation proposal \cite{K1},   the topological quantum gates, based on the MZMs and Majorana edge modes (MEMs) in the fractional quantum Hall effects, were designed \cite{Freedman2003,Geo,Geo1,TQCR,TQCR1}. 

Kitaev showed that MZMs can exist at the two ends of a  {topologically} superconducting quantum chain \cite{K3}. These MZMs are exact  {realizations} of those defined in \cite{NW,Inv,prox1}.   Using the proximity effect of the $s$-wave superconductor/topological insulator interface, Fu and Kane also presented an effective model to realize the MZMs \cite{fukane}. There are other proposals for MZMs, e.g., in semiconductor heterostructures \cite{Sau2010}, semiconductor-superconductor heterostructures \cite{Das Sarma2010}, and so on \cite{Hunting,theory,Das2013,Sau2015}. Experimentally, there  {have been} many reports on MZMs  {emerging}  exactly in pairs  \cite{Mourik2012,Deng2012,ADas,Chur,Deng,Nadj-Perge2014,Jia1,Sun2016}. Recently,  {evidence} of MZMs  {has been} reported on the surface of  iron-based superconductors, where the pairs of  MZMs are  {located} on different surfaces  {that} are separated by the topological bulk \cite{Kong2019,Chen2020,wang2020,Chen2019,P. Zhang2019,W. Liu2020}. 
Braiding these MZMs is the central focus of the Majorana qubit manipulation, and there are mainly two ways to braid the MZMs: the real space  or effective moving \cite{prox1,DM1,DM2,DM3,DM4,DM5,DM6,DM7} and  {the} measurement-only scheme \cite{Rau,Rau1,Measure1,Measure2,Measure3,boderson,Bravyi,measure CNOT}. The initial step in the former approach involves the experimental confirmation of the MZMs' existence, a task that continues to pose significant challenges. For the latter method, a comprehensive understanding   {of} it,  even for all Clifford gates, is not very well  {understood}. Recently,  {significant} progress  {has been made} in the designation of  measurement-only Majorana qubits \cite{LL}.  {For example,}  a measurement-based realization of the Pauli gates and the controlled-Z (CZ) gate without ancillary qubits  {has been achieved}. A new quantum computation process,  {based on fermion parity, is proposed as an alternative to the quantum circuit model \cite{Measure4}.}

The MEMs are  characteristic of chiral topological superfluidity  {and} superconductivity. Aside from that,  in the $\nu=5/2$ even denominator fractional quantum Hall state,   the chiral edge states of Kitaev's spin liquid \cite{K4}, the long-expected topological superconductor \cite{tosp1,tosp2}, and intrinsic  time-reversal breaking chiral superconducting states in a  {topologically} insulating thin film \cite{Luo1} are also regarded  {in} the MEMs.   In the $s$-wave superconductor/quantum anomalous Hall heterostructure,  the proximity effect may induce a chiral topological superconducting phase \cite{X.-L. Qi2010,BL}. Unfortunately, the experimental  {evidence} of this scenario does not seem likely nowadays \cite{X.-G. Wen2018,science1}. 
Thanks  {to} the propagating property of the MEMs, we can understand the Majorana-based quantum gates  {well} with the MEMs. Despite the experimental ambiguity, the theoretical studies for  topological quantum computation with the MEMs are convenient \cite{Freedman2003,TQCR,TQCR1,X.-L. Qi2010,BL,Geo,Geo1}. In our recent work \cite{Zhan},  {we used}  MEMs in  multilayers of the  chiral superconductor thin film \cite{hukane}  {to demonstrate} that these MEMs can  {function  as qubits for}  universal topological gates through their braidings. We also design  
quantum  {circuits} and quantum devices for quantum computing processes,  {such as} Shor's integer factorization algorithm \cite{Zhan}.

 {
There are two distinct encoding methods for Majorana-based topological quantum computation. In sparse encoding, $n$ logical qubits are realized using  $2^n$ physical qubits. This method aligns closely with the quantum circuit model and offers significant reusability for quantum gates. However, its primary limitation is the inability to generate entangled states through braiding alone, necessitating the use of numerous auxiliary bits combined with measurements to achieve such states. On the other hand, dense encoding constructs an $n$-logical qubit system with $2n+2$ physical qubits, inherently facilitating entanglement but encountering compatibility challenges with the tensor product structure inherent to the quantum circuit model. As the number of logical bits increases, a comprehensive redesign of all quantum gates becomes essential, negatively affecting the system's reusability and scalability. Considering  reusability and scalability, our recent work proposed a sparse-dense mixed encoding. Unlike 
solely relying on sparse encoding, this mixed encoding strategy eliminates the need for additional auxiliary qubits.}

However, it was recognized that  the braiding structures of some Majorana-based quantum gates are independent of the fermion parity (spin or charge) of the Majorana qubits.  {For example,} most of  {the} one-qubit gates and some direct product two-qubit gates.  {On the other hand}, some of them are dependent on the fermion parity \cite{Measure1,Measure2,Measure3,boderson,Bravyi,measure CNOT},  {such as} the controlled-NOT (CNOT) gate,  {which is a key component in constructing} the quantum circuit.  The common way to deal with the fermion parity dependence of the quantum gates is  {to make} a measurement with an ancillary qubit  {for} the computational state. When the measured state 
 {has} the desired fermion parity, the computational process continues,  {but when it does not have the desired parity}, this input is abandoned. Although it was estimated that such  {abandonment} is of acceptable efficiency in a real quantum computation process \cite{Zhan}, it still wastes many resources.

Instead of such a probabilistic process,  one may  {convert} the undesired fermion spin parity qubits to the desired one with some  {additional} costs \cite{Bravyi,measure CNOT}. However, due to the use of sparse encoding \cite{sparse}, implementing  the fermion spin parity correction  {depends} on two measurements. This causes the complexity  {of} the computational process, as we will describe  {later}. Furthermore, this correction operation cannot be applied to the sparse-dense mixed encoding  {in} which we encode the qubits in \cite{Zhan}.  
The authors of Ref. \cite{LL} present  a  simpler operation  {that is relevant to converting}  the undesired outcome  {into} the desired one for  measurement-only topological quantum computation. It is argued that no ancillary qubit is required in the correction operation. These authors consider a hybrid of superconductor/two-dimensional topological insulator/ferromagnetic insulator  where MZMs lie and  {propose} the deterministic Clifford gates.      

In this work, we will study a correction operation  {for} the fermion charge parity  {in} the sparse-dense mixed encoding quantum computing process. We aim to improve our probabilistic quantum process in  \cite{Zhan} to  {a} deterministic one through  {an} efficient fermion charge parity correction operation. In the sparse encoding sense, we also show that no additional ancillary qubit is required.  Another way to resolve  {the problem of mismatched parity} between the input qubits and the quantum gates is to correct the quantum gates,  {as} briefly described in our previous work \cite{Zhan}.  {Here,} we  {provide the specific}  process. By  {incorporating} the correction process, we  {can achieve} a dissipationless universal topological quantum computer.
We discuss the efficiency of our process. Comparing with the correction  process in \cite{Bravyi,measure CNOT}, our  process is more efficient.  

This work is organized as follows:  In Sec. II, we briefly summarize the main results  {of} our previous work based on the sparse-dense encoding process \cite{Zhan}. We also review the fermion parity correction procedures provided in Ref. \cite{measure CNOT,LL} for the sparse encoding.  In Sec. III,
we discuss the fermion charge parity correction for the sparse-dense mixed encoding in terms of our proposed quantum gates. The example for the dissipationless CNOT gate with the MEMs is explicitly implemented, in terms of the multilayers of the chiral  superconductor thin film \cite{hukane, Zhan}. The efficiency of our computational process  {is} estimated and compared with  {that of} other processes.

\section{Majorana qubits,  fermion parity, and Topological Quantum gates }

\subsection{The basis}

The quantum  {state} space of the multi-Majorana objects  {is} identical to the quantum states of the quantum Ising model \cite{NW, Inv}.  According to Ivanov \cite{Inv},  Majorana fermions appear in pairs $(\gamma_1,\gamma_2)$ with a phase difference  $\pi/4$, namely, when they fuse, $\gamma_1^2=\gamma_2^2=1$ but up to an overall phase  $(\gamma_1,\gamma_2)\to \Psi_A\propto (\gamma_{A1}+i\gamma_{A2})/2$ or $\Psi_A^\dag \propto (\gamma_{A1}-i\gamma_{A2})/2$, i.e., annihilating or creating a conventional fermion  {labeled} as $A$ with 
$\{\Psi_A,\Psi_A^\dag\}=1, \{\Psi_A,\Psi_A\}=\{\Psi_A^\dag,\Psi_A^\dag\}=0$.  The Pauli gates are $X_A,Y_A, Z_A$ acting on the one-qubit for the basis $(|0_A\rangle, \Psi_A^\dag |0_A\rangle)^T\equiv (|0_A\rangle, |1_A\rangle)^T$, called  {the} Majorana one-qubit.    The Majorana two qubit gates are then defined  {in} the basis of $(|00\rangle, \Psi_A^\dag |00\rangle,\Psi_B^\dag|00\rangle, \Psi_A^\dag\Psi_B^\dag |0\rangle)^T\equiv(|0_A0_B\rangle, |1_A0_B\rangle,|0_A1_B\rangle,  |1_A1_B\rangle)^T $.  The fermion charge parity of a quantum state is defined by $(-1)^F$, where $F$ is the fermion number of the state. The fermion charge parity of the state $|0\rangle$ or $|1\rangle$ is $\pm1$, while the states  $(|0_A0_B\rangle, |1_A0_B\rangle,|0_A1_B\rangle,  |1_A1_B\rangle)^T $ have $(-1)^F$ to be $(+,-,-,+)$. Notice that such a notion can also be applied to the spin system if identifying $|\downarrow\rangle\equiv|0\rangle$ and $|\uparrow\rangle\equiv|1\rangle$. In this case, the charge parity is replaced by the spin parity \cite{measure CNOT}.

 For a particle number conserved system, the quantum state is characterized by $F$ but not $(-1)^F$.  But when fermions are created and annihilated in pairs, such as  in superconductors, the conserved quantity is $(-1)^F$. In these systems,  the Majorana one-qubit under the basis $(|0_A\rangle, |1_A\rangle)$ and Majorana two-qubits in the basis $(|0_A0_B\rangle, |1_A0_B\rangle,|0_A1_B\rangle,  |1_A1_B\rangle)^T$  are meaningless because of the fermion parity conservation. The minimal basis for  a one-qubit is then $|\pm_{AB}\rangle_1$, where $|+\rangle_1=(|0_A0_B\rangle,|1_A1_B\rangle) ^T$ and $|-\rangle_1=(|0_A1_B\rangle,|1_A0_B\rangle)^T$ with $\pm$ referring to the fermion parity of the states.  We have seen that  most one-qubit gates are independent of the fermion parity of the basis (e.g., see \cite{Zhan}). 
 
 For the two-qubit gates, the dense encoding  process takes the minimal basis with three pairs of Majorana objects \cite{Freedman2003,TQCR,TQCR1} 
 \begin{eqnarray}
 &|+_{ABC}\rangle_{2}=(|0_A0_B0_C\rangle,|0_A1_B1_C\rangle,|1_A0_B1_C\rangle,|1_A1_B0_C\rangle)^T&\nonumber\\
 &|-_{ABC}\rangle_{2}=(|0_A0_B1_C\rangle,|0_A1_B0_C\rangle,|1_A0_B0_C\rangle,|1_A1_B1_C\rangle)^T.&\nonumber
 \end{eqnarray}
 The two-qubits with the dense encoding  {are} abbreviated as two-Dqubits.
 
 \subsection{One-qubit gates}
 
{There are two important kinds of elementary mutual braidings in the construction of the one-qubit gates in the fermion parity basis of Majorana objects. One is the braiding of Majorana fermions within a given fermion, say, $R_{12}$ ($\gamma_{A1}\leftrightarrow \gamma_{A2}$), and the other one is the braiding of Majorana fermions between two different fermions, say $B_{23}$ ($\gamma_{A2}\leftrightarrow \gamma_{B3}$). For MZMs, Ivanov \cite{Inv} showed that in both 
 {parity} even basis $|+\rangle_1$ and parity odd basis $|-\rangle_1$, 
 \begin{equation}
     R_{12}=\left(
		\begin{array}{ccc}
		e^{-i\pi/4}  & 0 \\
		0 &  e^{i\pi/4}
		\end{array}
		\right),\quad
		B_{23}=\frac{1}{\sqrt{2}}\left(
		\begin{array}{ccc}
		1  & -i \\
		-i &  1
		\end{array}
		\right),
 \end{equation}
 whereas the manipulation of moving the MZMs remains an experimental task.}
 
In the following, we recall our one-qubit gates with the MEMs \cite{X.-L. Qi2010,BL,Zhan}.  We describe the exchanging,  braiding, and entangling of the MEMs in the parity basis $|\pm_{AB}\rangle_1$. Correspondingly, they are the exchanging, braiding, and fusion matrices in the unitary modular tensor category.  In our scenario, the exchange operations act on $\gamma_{1A},\gamma_{2A}$ which are two MEMs from the same fermion $\Psi_{A}$ and similarly for $\gamma_{B3},\gamma_{B4}$. It was found that in a seven- {layered} effective chiral  superconductor, the MEMs $\gamma_{1,2}$ can be decomposed into a pair of Fibonacci  {anyons} $\tau$ with conformal   {dimensions of} $2/5$ and a pair of $\varepsilon$  {anyons} with conformal  {dimensions of} $1/10$ \cite{hukane}. By denoting $G(\theta)=e^{i2\theta}$, then, through exchanging  $\varepsilon_1,\varepsilon_2$ (or $\tau_1,\tau_2$) twice in $\gamma_1$ and $\gamma_2$, both of $\gamma_1$ and $\gamma_2$  {obtain} a phase factor $G(\frac{\pi}{10})=e^{i\frac{\pi}5}$  (or $G(\frac{2\pi}5)=e^{i\frac{4\pi}5}$) while exchanging each pair $\varepsilon_{1,2}$ and $\tau_{1,2}$ once, they obtain a phase factor $G(-\frac{\pi}4)$ \cite{Zhan}.   According to these phase factors, we can have the one-qubit phase gates $R_{12}(\theta/2)={\rm diag}(1, G(-\theta))$. Physically, this phase gate means that if the fermion number $n_A=0$, nothing happens, and if $n_A=1$, the fermion gains a phase $G(-\theta)$. It is easy to see  {that} these phase gates are independent of the fermion parity of the basis. However, we find that $R_{34}^{(-)}(-\frac{\pi}4)$ is not the same as $R_{34}^{(+)}(-\frac\pi{4})$ because $R_{34}^{(+)}(-\frac\pi{4})={\rm diag}(1,i)$ while $R_{34}^{(-)}(-\frac\pi{4})={\rm diag}(i,1)$.  
{These parity-dependent results are consistent with those in Ivanov \cite{Inv}, and they only differ by an overall constant phase. This parity dependency and its influence on designing the Majorana-based quantum device  {is} not well considered in the literature, which is in fact the  {reason} why multi-qubit gates are dependent on the parity of  {the} basis, and  {a} parity correction process is needed. Later, we will use the parity dependence of CNOT gate as an example and discuss the corresponding correction process. }

The braiding gate $B_{23}$ for $\gamma_{2A}$ and $\gamma_{3B}$ under the basis $|\pm_{AB}\rangle_1$ for the MEMs has been proposed in \cite{X.-L. Qi2010,BL}, which is not dependent on the fermion parity and is given by 
 \begin{equation}
		B_{23}=\frac{1}{\sqrt{2}}\left(
		\begin{array}{ccc}
		i  & 1 \\
		1 &  i
		\end{array}
		\right). \label{B23}
		\end{equation}
{This braiding process can be realized by the scattering of MEMs between a metal and a chiral topological superconductor, which is also topological \cite{X.-L. Qi2010,BL,Zhan} and easier than that of MZMs.} And the Hadamard gate is given by 
\begin{eqnarray}\label{H}
R_{12}(-\frac{\pi}4)B_{23}R_{12}(-\frac{\pi}4)=\frac{i}{\sqrt2}\left(
		\begin{array}{ccc}
		1 & 1 \\
		1 & -1
		\end{array}
		\right)=iH.
	\end{eqnarray}
With $H^{-1}=H$, one has $$B_{23}=e^{-i\frac{\pi}4}HR_{12}(-\frac{\pi}4)H^{-1}.$$ This is the duality relation between fusion, exchange, and braiding in the unitary modular tensor category, $B=F^{-1}R F$, where $F=H^{-1}=H$ and $R=e^{-\frac{\pi}4}R_{12}$ \cite{MS}. Furthermore, we have $ Z=(R_{12}(-\frac{\pi}4))^2,Y=(HZ)^2$ and 
\begin{eqnarray}
X=-iB_{23}^2=-H(R_{12}(-\frac{\pi}4))^2H.\label{dual}
\end{eqnarray} 
The NOT gate $X$ is given by braiding the MEM from $\Psi_A$ and the MEM from $\Psi_B$ twice. It is easy to check that $H$, $R_{12}(-\frac{\pi}4)$, and $B_{23}$ obey the pentagon and hexagon equations in the  unitary modular tensor category \cite{MS}.  

Notice that $R_{12}(-\frac\pi{10})$ and $R_{12}(-\frac{2\pi}5)$ do not belong to the Clifford gates \cite{Zhan}. Then, the quantum gates  $\{H,R_{12}(-\frac{\pi}4),R_{12}(-\frac\pi{10}), CNOT\}$ form a universal set, and any elements of  $SU(2)$ can be realized  {with} desired precision \cite{Zhan,univ}.   We have also designed the corresponding devices  {for} these gates with the multilayers of the chiral  superconductor thin film \cite{Zhan}. For example, we recall the device designations for the Pauli gates $X,Y,Z$ in Appendix A because they will play important roles in the fermion parity correction process.

 \subsection{Qubits and parity corrections in sparse encoding}
 
 In the sparse encoding process \cite{sparse},  four pairs of Majorana objects are considered for a two-qubit gate. Since there are eight basis states in a given fermion parity subspace, one has to choose four of them to span the computational space. We abbreviate such two-qubits as two-Squbits.   In the computational process,  because the mixing between the computational states and the non-computational states cannot be avoided,  the process either becomes probabilistic or needs to be corrected.  As the  {cost} of the fermion parity correction, an additional ancillary qubit, i.e., four Majorana objects,  {is} introduced \cite{Bravyi,measure CNOT}.  The advantage of the correction process in \cite{measure CNOT} is that, instead of the long-range braiding operation \cite{Bravyi}, only the local braidings are needed. On the other hand, besides introducing ancillary  {qubits}, for the control-target two-qubit gates, both control and target qubits need to be corrected according to two measurements.  
 
 
  {In the measurement-only process, Zhang $et$ $al$. \cite{LL} proposed a scheme to correct the undesired outcome that arose from the computational operation recently. In this model, the operations of fusion and braiding are effectively conducted through measurements rather than physically relocating Majorana fermions. However, these measurement processes can lead to the emergence of undesired quantum states.}
 For example, they take the computational Squbit spanned by the basis
 \begin{eqnarray}
 (|0_A0_B0_C0_D\rangle,  |0_A0_B1_C1_D\rangle,|1_A1_B0_C0_D\rangle,|1_A1_B1_C1_D\rangle)^T.\label{e41}
 \end{eqnarray} 
 The non- {computational} basis, but with the same fermion parity, may be mixed with the computational basis due to the computational operations, which is
 \begin{eqnarray}
 (|0_A1_B0_C1_D\rangle,  |0_A1_B1_C0_D\rangle,|1_A0_B0_C1_D\rangle,|1_A0_B1_C0_D\rangle)^T.\label{e42}
 \end{eqnarray}
 Zhang $et$ $al$. found that if using the parity correction operation to the Majorana one-qubit $(|0_{B,C}\rangle, |1_{B,C}\rangle)$, i.e., $|0_{B,C}\rangle\leftrightarrow|1_{B,C}\rangle$, the non-computational basis could be turned back  {to} the computational one.  {Therefore, the required branding operation can be completed correctly and deterministically.} They further showed that without adding an ancillary one-qubit, the correction operation could be realized by turning on the effective $X_{B,C}$ gates when the undesired parity is measured for the states at $B,C$ according to the duality relation between the fusion, exchange, and braiding, that is, $X_{B,C}=F^{-1}_{B,C}R^2_{B,C} F_{B,C}$, similar to  Eq. \eqref{dual}.

\subsection{Two-qubit gates in dense encoding}
 
 We  recall the CNOT gate in the dense encoding \cite{Geo1,Zhan}. We define $\Psi_A=(\gamma_{A1}+i\gamma_{A2})/2$,  $\Psi_B=(\gamma_{B3}+i\gamma_{B4})/2$ and $\Psi_C=(\gamma_{C5}+i\gamma_{C6})/2$. Under the basis $|+_{ABC}\rangle_2$, the CNOT is given by 
 \begin{eqnarray}\label{cnot+}
&&{\rm CNOT}^{(+)}=B_{45}^{(2)}R^{(2)}_{34}(-\frac{\pi}4)R^{(2)}_{56}(-\frac{\pi}4)B^{(2)}_{45}\label{CNOT+}\\
&&R^{(2)}_{56}(-\frac{\pi}4)R^{(2)}_{34}(-\frac{\pi}4)R^{(2)-1}_{12}(-\frac{\pi}4)=\left(
		\begin{array}{cccc}
		1  & 0 & 0 & 0 \\
		0  & 1 & 0 & 0 \\
		0  & 0 & 0 & 1 \\
		0  & 0 & 1 & 0
		\end{array}
		\right),~~~\nonumber
\end{eqnarray}
where the two-Dqubit gates $R_{12}^{(2)}(-\frac{\pi}4)={\rm diag}(1,1,i,i)$, $R_{34}^{(2)}(-\frac{\pi}4)={\rm diag}(1,i,1,i)$, and  $R_{56}^{(2)}(-\frac{\pi}4)={\rm diag}(1,i,i,1)$. $B_{45}^{(2)}$, as the counterpart of $B_{23}$,  is given by
\begin{eqnarray}
		B_{45}^{(2)}=
		\frac{1}{\sqrt{2}}\left(
		\begin{array}{cccc}
		i  & 1 & 0 & 0 \\
		1  & i & 0 & 0 \\
		0  & 0 & i & 1 \\
		0 & 0 & 1 & i
		\end{array}
		\right).\nonumber
		\end{eqnarray}
However, it is easy to see  {that} the CNOT gate is dependent on the fermion parity. If we design the device according to the two-Dqubit gate sequence given by \eqref{cnot+}. 
The result is not the right side of \eqref{cnot+} \cite{Zhan}. The correct two-Dqubit gate sequence, which acts on $|-_{ABC}\rangle_2$ and gives the CNOT$^{(-)}$  {is} as follows	
	 \begin{eqnarray}\label{cnot-}
&&{\rm CNOT}^{(-)}=B_{45}^{(2)}R^{(2)-1}_{34}(-\frac{\pi}4)R^{(2)}_{56}(-\frac{\pi}4)B^{(2)}_{45}\nonumber\\
&&R^{(2)-1}_{56}(-\frac{\pi}4)R^{(2)}_{34}(-\frac{\pi}4)R^{(2)}_{12}(-\frac{\pi}4).\label{CNOT-}
\end{eqnarray}

 \subsection{Qubits and universal topological quantum gates in sparse-dense mixed encoding process} 		

With three gates $\{H, Z^{\frac{1}{4}}, {\rm CNOT}\}$,  the quantum circuit models for quantum computation can  {be} built \cite{univ, UTQC}.
To prove universality, the key essence is to create two orthogonal axes and a phase with  {an} irrational number times $\pi$ from the universal set, namely, any element in $SU(2)$ can be approximated  {with the} desired precision. {As mentioned before, the set of gates $\{H,\sqrt{Z}, R(-\frac{\pi}{10}),{\rm CNOT}\}$  
is universal \cite{Zhan}. Therefore}, in principle, we can design the universal topological quantum computer. However, the dense encoding process is not directly relevant to a quantum circuit model. Therefore,  we use the sparse-dense encoding process, i.e., taking the sparse encoding  process to fit the quantum circuit model while processing the quantum computing by the dense encoding. 

\begin{figure}[!h]
        \centering \includegraphics[width=1\columnwidth]{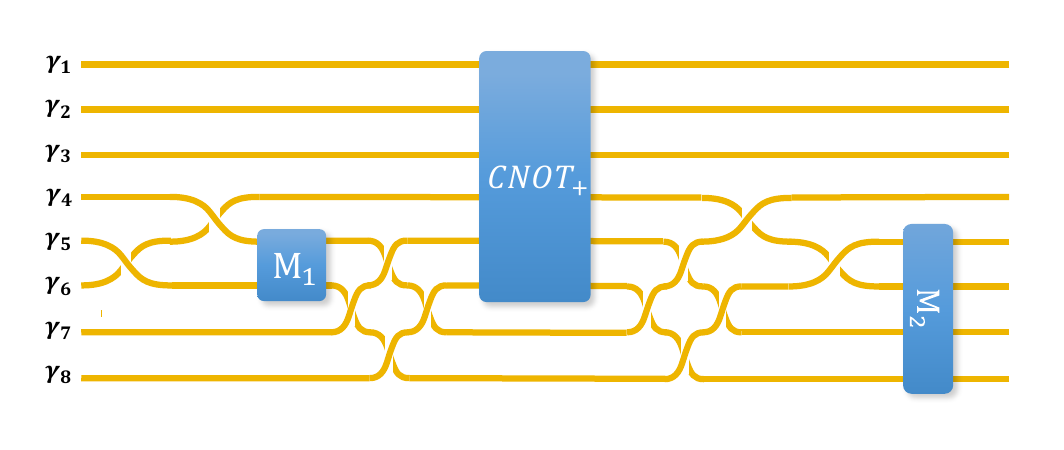}
        \caption{(Color online) The CNOT  {is} in sparse encoding with even parity. The blue  {shaded} area is the CNOT$^{(+)}$ in the dense encoding. 
        }\label{fig1}
\end{figure}

In our previous work \cite{Zhan}, we  {started} from the sparse encoding basis. To form a quantum circuit, we require each one-qubit  {to have} the same fermion parity, say, $+$. Then two-Squbits are of the even fermion parity with the basis given by  \eqref{e41} and \eqref{e42}, denoted  as $|+_{ABCD}\rangle_4$. The two-qubits are formed by four pairs of MEMs: $(\gamma_{A1},\gamma_{A2}), (\gamma_{B3},\gamma_{B4}),(\gamma_{C5},\gamma_{C6}),(\gamma_{D7},\gamma_{D8})$, 
and the initial input is enforced by $ \gamma_{A1}\gamma_{A2}\gamma_{B3}\gamma_{B4}=-1$ and $\gamma_{C5}\gamma_{C6}\gamma_{D7}\gamma_{D8}=-1$. To reduce to dense encoding, we take $(\gamma_{4B},\gamma_{5C})$ as the ancillary Majorana one-qubit. Before making the measurement $M_1$, we entangle  the first qubit and the second qubit with a series of  {exchanges} and braiding, as shown in Fig. \ref{fig1}. If the measurement $M_1$ gives  {a} positive fermion parity, then the rest of  {the} MEMs propagate through
the blue  {shaded} area, which is a CNOT$^{(+)}$ in the dense encoding. The final step is measuring the output state. If $M_2$ gives    $\gamma_{C5}\gamma_{C6}\gamma_{D7}\gamma_{D8}=-1$, this output state is the desired state ready for the next computational process.  This is our quantum process for sparse-dense encoding. 

The controlled-Y(CY) and CZ gates are obtained in terms of CNOT
\begin{eqnarray}
{\rm CY}=R_{12}(-\frac{\pi}4){\rm CNOT}~ R_{12}^{-1} (-\frac{\pi}4),~  {\rm C}(i{\rm Z})= {\rm CNOT\cdot CY}.
\end{eqnarray}
This means that these two-Dqubit gates are also dependent on the fermion parity.

\section{The fermion parity correction for the sparse-dense encoding process}

As we have seen in the sparse-dense mixed encoding process, the input information will be abandoned when the undesired fermion parity is detected either in $M_1$ or $M_2$. This makes the topological quantum computing process probabilistic. Although it has been estimated that this probabilistic process  {is} still of  good efficiency \cite{Zhan},  vast quantum information resources are wasted. Similar to that in the sparse encoding process \cite{measure CNOT, LL},  it is better to introduce the correction when the undesired fermion parity is detected. Two ways of the correction processes will be presented below, {and they work for both MZM-based and MEM-based quantum computations.}

\subsection{Process I: correcting the input qubits }

   As we have said, at a given $A$, the operation $X_A$ switches the basis $|0_A\rangle\leftrightarrow|1_A\rangle$ and then exchanges the fermion parity of the basis. However, in the basis we take, there is no $X_A$ operation. In our basis, the one-qubit space is spanned by  $(|+\rangle_1)=(|0_A0_B\rangle,|1_A1_B\rangle)^T$ and $(|-\rangle_1)=(|0_A1_B\rangle,|1_A0_B\rangle)^T$. The $X=-iB_{23}^2$ gate does not change the fermion parity of the basis because $X(|0_A0_B\rangle,|1_A1_B\rangle)^T=(|1_A1_B\rangle,|0_A0_B\rangle)^T$ and  $X(|0_A1_B\rangle,|1_A0_B\rangle)^T=(|1_A0_B\rangle,|0_A1_B\rangle)^T$. However, for  {a} given $A$ and $B$, we see that $|1_A\rangle\leftrightarrow|0_A\rangle$ and $|1_B\rangle\leftrightarrow|0_B\rangle$, i.e., $X=X_A\otimes X_B$.  
    
\begin{figure}[!h]
        \centering \includegraphics[width=1\columnwidth]{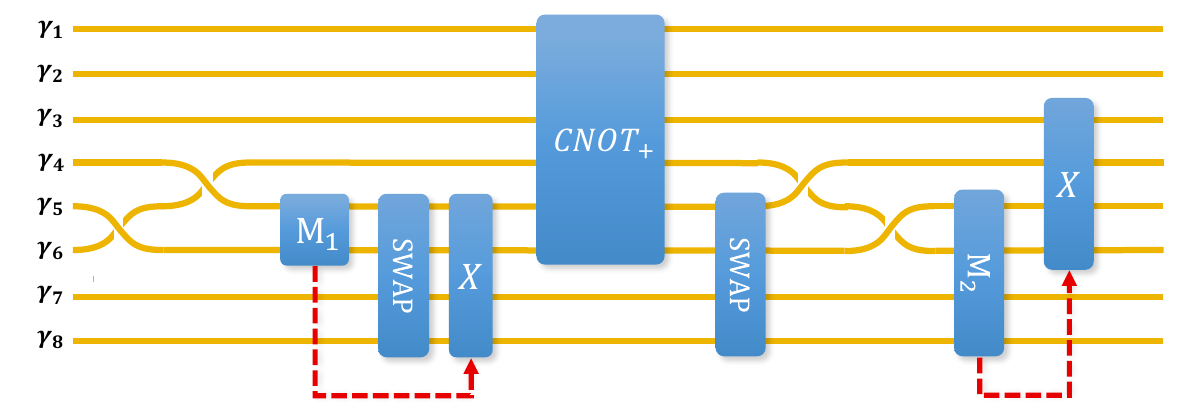}
        \caption{(Color online) Correcting the undesired fermion parity measured. $M_1$ and $M_2$ for the CNOT gate in the sparse-mixed encoding process. 
        }\label{fig2}       
\end{figure}

We now show the details of the correction process  {for} the CNOT gate. We have  {labeled} 4 pairs of $(\gamma_1,\gamma_2),(\gamma_3,\gamma_4), (\gamma_5,\gamma_6),(\gamma_7,\gamma_8)$  as $A,B,C,D$. We take $\gamma_{4,5}$ as an ancillary qubit and  measure its fermion parity, i.e., $i{\gamma}_{4}{\gamma}_{5}$. We label this measurement as $M_1$ in {Fig. \ref{fig2}}.  If the measurement $M_1$  results in a positive fermion parity, we do nothing. For a negative fermion parity,  i.e., an electron $\Psi^\dag_C=(\gamma_{C4}-i\gamma_{C5})/2$ is detected, the fermion parity correction can be operated as follows:  

Since the total fermion parity is conserved, this means that the left two-Dqubits constituted by $\gamma_1,\gamma_2,\gamma_3,\gamma_6,\gamma_7,\gamma_8$  {are} also of the negative fermion parity.  To correct this undesired fermion parity, 
let this electric signal switch on the gate $X$ acting on the one-qubit constituted by $\gamma_{C4},\gamma_{C5},\gamma_{D7},\gamma_{D8}$. This  {corresponds} to braiding $\gamma_{C8}$ and $\gamma_{D4}$ twice [see {Fig. \ref{fig2}}]. Since $X=X_C\otimes X_D$, the fermion parity $i\gamma_{C4}\gamma_{C5}$ changes from $-$ to $+$ while $i\gamma_{D7}\gamma_{D8}$ also changes sign, and so $\gamma_1\gamma_2\gamma_3\gamma_6\gamma_7\gamma_8$ changes from $-$ to $+$. Therefore, the undesired fermion parity of the computational state in the dense encoding is corrected to the desired one.  

We now describe how to make the measurement $M_1$.   If we choose the even fermion parity input state given by ${\gamma}_{1}{\gamma}_{2}{\gamma}_{3}{\gamma}_{4}={\gamma}_{5}{\gamma}_{6}{\gamma}_{7}{\gamma}_{8}=-1 $ \cite{sparse}, the corresponding computational two-Squbit basis is given by
\begin{eqnarray}
(|0_A0_B0_C0_D\rangle,|0_A0_B1_C1_D\rangle,|1_A1_B0_C0_D\rangle,|1_A1_B1_C1_D\rangle)^T.\label{evenB}
\end{eqnarray}
At this stage,  {measuring} $i\gamma_4\gamma_5$ does not make sense because $\gamma_4$ and $\gamma_5$ do not form a fermion associated with the above basis. 
 To be the fermion associated with that in the basis, let $\gamma_{6}$ exchange with $\gamma_{4,5}$ and form a new pair with $\gamma_{3}$, and $(\gamma_4,\gamma_5)$ form another new pair [see, {Fig. \ref{fig2}}]. This  {is equivalent} to  {applying} $B_{45}B_{56}$  {to} the qubit $ \{({\gamma}_{3},{\gamma}_{4}),({\gamma}_5,{\gamma}_{6})\} $ and  {results in} a superposition of the computational and non-computational two-Sqbasis 
 \begin{equation}
\left(
\begin{array}{ccc}
i|0_A0_B0_C0_D\rangle+|0_A1_B1_C0_D\rangle \\
-|0_A0_B1_C1_D\rangle+i|0_A1_B0_C1_D\rangle \\
|1_A0_B1_C0_D\rangle+i|1_A1_B0_C0_D\rangle \\
i|1_A0_B0_C1_D\rangle-|1_A1_B1_C1_D\rangle \\
\end{array}
\right), \label{SP}
\end{equation}
where $A,B,C,D$ label the Majorana fermion pairs $(\gamma_1,\gamma_2),(\gamma_3,\gamma_6), (\gamma_4,\gamma_5),(\gamma_7,\gamma_8)$.

We then measure $i{\gamma}_{4}{\gamma}_{5}$, i.e.,  {performing } $M_1$ in {Fig. \ref{fig2}}. When it is $+1$, the fermion number of the pair $C$ is 0, and the basis \eqref{SP} collapses to a two-Dqubit basis
\begin{eqnarray}
(|0_A0_B0_D\rangle,|0_A1_B1_D\rangle,|1_A1_B0_D\rangle,|1_A0_B1_D)^T,\label{Dbasis}
\end{eqnarray}
which is the basis of the 2-qubits for the dense encoding in the even fermion parity. 

 When  $i{\gamma}_{4}{\gamma}_{5}=-1$, the fermion number of the pair $C$ is 1, and the collapsed two-Dqubit basis is given by 
\begin{eqnarray}
(|0_A1_B1_C0_D\rangle,-|0_A0_B1_C1_D\rangle,|1_A0_B1_C0_D\rangle,-|1_A1_B1_C1_D\rangle)^T.
\label{swap}
\end{eqnarray}

 Here we introduce a quantum gate called the SWAP, which has the ability to exchange fermionic numbers between sites C and D, and is represented as follows:
 \begin{eqnarray}
		SWAP=
		\frac{1}{\sqrt{2}}\left(
		\begin{array}{cccc}
		1  & 0 & 0 & 0 \\
		0  & 0 & 1 & 0 \\
		0  & -1 & 0 & 0 \\
		0 & 0 & 0 & 1
		\end{array}
		\right).\nonumber
		\end{eqnarray}
{The braiding diagram for the SWAP operation is shown in Appendix C. Applying the SWAP} to CD {of (\ref{swap})},
\begin{eqnarray}
(|0_A1_B1_C0_D\rangle,|0_A0_B1_C1_D\rangle, |1_A0_B1_C0_D\rangle,|1_A1_B1_C1_D\rangle)^T.
\end{eqnarray}
Notice that the overall phase of -1 has been omitted. Acting $X_C\otimes X_D$ on this remaining basis,  it becomes
 \begin{eqnarray}
(|0_A1_B0_C1_D\rangle,|0_A0_B0_C0_D\rangle, |1_A0_B0_C1_D\rangle,|1_A1_B0_C0_D\rangle)^T.\label{2dq}
\end{eqnarray}

 As we have  said, this correction can be  realized by braiding $\gamma_5$ and $\gamma_7$ twice, i.e., $B_{57}^2\propto X=X_C\otimes X_D$, up to a total phase factor [see {Fig. \ref{fig2}}].  This operation also switches the pair $(\gamma_4,\gamma_5)$  from $i\gamma_4\gamma_5=-1$ to $1$.  
 In the case of even parity, we also exchange the positions of C and D by employing the SWAP gate, {and no additional phase is introduced}.

 The corrective approach is not unique.{The basis \eqref{2dq}} and that in \eqref{Dbasis} only differ by a linear transformation, which can be realized using a two-qubits gate.
 Another realization process and the definitions,  {and} implementations of the SWAP, SWAP$'$, and $X$ gates are provided in the Appendix for reference.


Continuing the process and  {applying} the CNOT$^{(+)}$  {gate}, on the desired basis  state \eqref{Dbasis}, the output basis is given by
$$(|0_A0_B0_D\rangle,|0_A1_B1_D\rangle,|1_A0_B1_D\rangle,|1_A1_B0_D\rangle)^T$$
Putting the pair $(\gamma_4,\gamma_5)$ with $i{\gamma}_{4}{\gamma}_{5}=1$ back, we have
$$(|0_A0_B0_C0_D\rangle,|0_A1_B0_C1_D\rangle,|1_A0_B0_C1_D\rangle,|1_A1_B0_C0_D\rangle)^T.$$
To complete this sparse-dense encoding process, we want to restore the input basis, i.e.,  {perform} the operations  before the measurement $M_2$. This means that we need to
braid $\gamma_{6}$ with ${\gamma}_{4}$ and ${\gamma}_{5}$ again, and  we then  {obtain} the superposition state with $\gamma_1\gamma_2\gamma_3\gamma_4$  (and  $\gamma_5\gamma_6\gamma_7\gamma_8$) either $=+1$ or $=-1$, i.e.,
\begin{equation}
\left(
\begin{array}{ccc}
|0_A0_B0_C0_D\rangle+|0_A1_B1_C0_D\rangle \\
|0_A0_B1_C1_D\rangle+|0_A1_B0_C1_D\rangle \\
|1_A0_B0_C1_D\rangle+|1_A1_B1_C1_D\rangle \\
|1_A0_B1_C0_D\rangle+|1_A1_B0_C0_D\rangle \\
\end{array}
\right).
\end{equation}
We then  measure  ${\gamma}_{5}{\gamma}_{6}{\gamma}_{7}{\gamma}_{8}$, i.e.,  the measurement $M_2$ in Fig.\ref{fig2} (a). If it is $-1$, we have the fermion parity even output basis in the sparse encoding
$$(|0_A0_B0_C0_D\rangle,|0_A0_B1_C1_D\rangle,|1_A1_B1_C1_D\rangle,|1_A1_B0_C0_D\rangle)^T.$$
 If the measurement $M_2$ is odd, we have the fermion parity odd output basis in the sparse encoding, 
$$(|0_A1_B1_C0_D\rangle,|0_A1_B0_C1_D\rangle,|1_A0_B0_C1_D\rangle,|1_A0_B1_C0_D\rangle)^T.$$
We then act $X=X_B\otimes X_C$ on the above output [see {Fig. \ref{fig2}}]. The basis goes back  {to} the  computational basis \eqref{evenB}.


 {This approach adjusts the fermion parity from an odd output basis to an even output basis, subsequently facilitating the implementation of a CNOT gate within sparse encoding. Without performing correction operations and discarding undesired states, the  probability of achieving the correct outcome after $N$ measurements is $\frac{1}{2^N}$. However, through corrections with a polynomial time complexity of $O(N)$, undesired states are transformed into desired states. This modification ensures that each result is deterministic, avoiding the exponentially decreased probability associated with uncorrected processes. It's noteworthy that, in contrast to the sparse encoding process described in \cite{measure CNOT}, our method eliminates the need for additional ancillary qubits, a realization also achieved by \cite{LL} through a measurement-only scheme.}

Furthermore, according to the above sparse-dense encoding process, any parity dependent two-Dqubit gate can be generalized to the two-Squbit gate in the same way because our process in fact does not specify the two-Dqubit gate. For a parity independent two-Dquibt gate, the  fermion parity correction can be omitted after the measurement $M_1$ [see  {Fig. \ref{fig6}} in the Appendix].

Therefore, with  the universal set of gates $\{H,\sqrt{Z}, R(-\frac{\pi}{10}),{\rm CNOT}\}$, we can perform dissipationless universal topological quantum computation. In  {terms of}  sparse encoding, no additional ancillary qubit is needed for redundant measurements.    

\subsection{Process II: correcting the gates in the dense encoding}
 
\begin{figure}[!h]
        \centering \includegraphics[width=1\columnwidth]{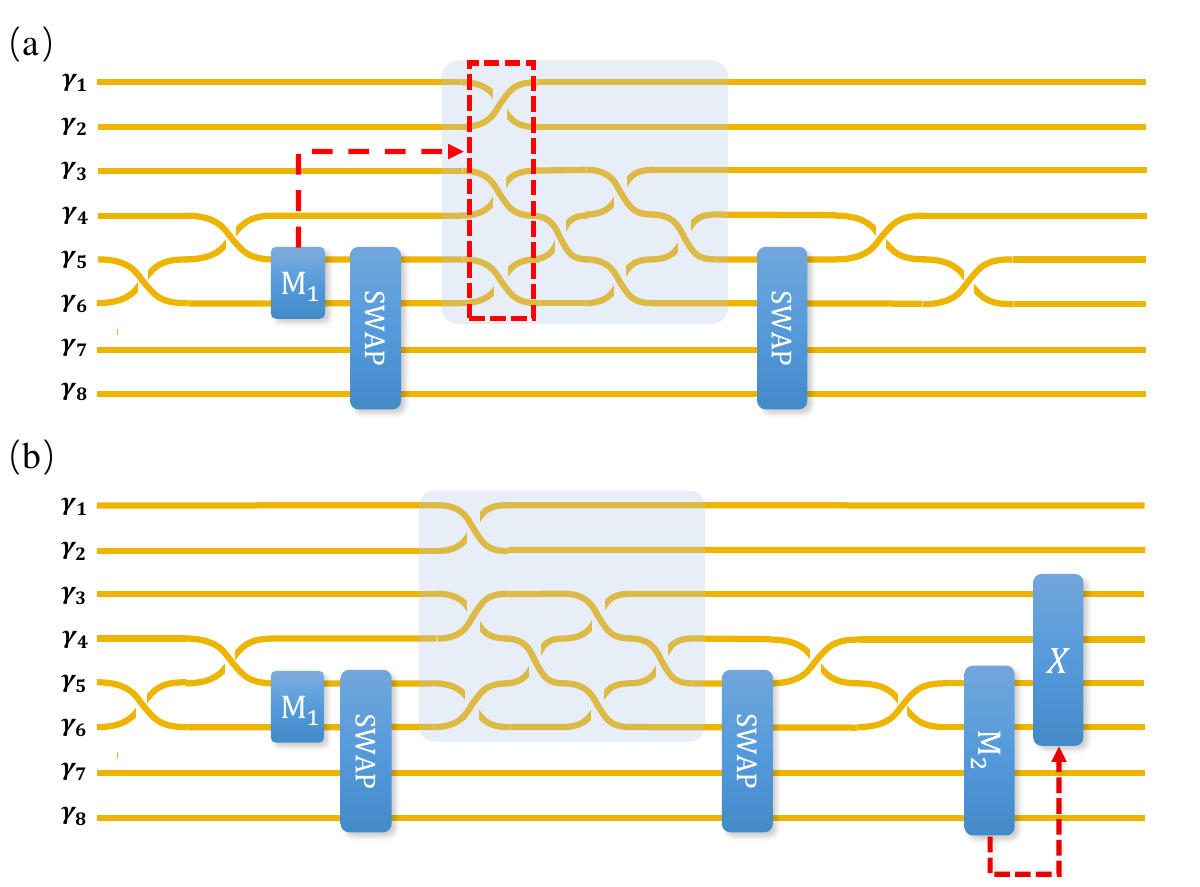}
        \caption{(Color online) Correcting the fermion-dependent gates. {The blue transparent shade corresponds to the CNOT gate.} (a) before correction; (b) after correction.
        }\label{fig3}       
\end{figure}

As we mentioned in the previous work \cite{Zhan},  we can also try to correct the parity-dependent gate if the measurement is undesired, instead of correcting the input data. For example, we start from the input as  {described} in the last subsection. The braiding data of the CNOT gate in the sparse encoding is shown in Fig. \ref{fig3} (a) where the blue shade is the data of the CNOT$^{(+)}$ for the dense encoding. We first do the measurement $M_1$. If the result is positive fermion parity, we do nothing as before. If the negative parity is detected,  three signals are sent to three phase gates [see Fig. \ref{fig3} (a)].   
These signals give orders so that the phase gates $R^{(2)-1}_{12}(-\frac{\pi}4), R^{(2)}_{34}(-\frac{\pi}4),R^{(2)}_{56}(-\frac{\pi}4)$ turn to $R^{(2)}_{12}(-\frac{\pi}4), R^{(2)-1}_{34}(-\frac{\pi}4),R^{(2)-1}_{56}(-\frac{\pi}4)$ and the CNOT$^{(+)}\to$ CNOT$^{(-)}$ [see Fig. \ref{fig3} (b)]. According to \eqref{CNOT+} and \eqref{CNOT-}, this corrects the CNOT$^{(+)}$ to the CNOT$^{(-)}$. In practice,  since $G(\mp\frac{\pi}4)= G(\pm\frac{\pi}4) G(\pm\frac{\pi}4)G(\pm\frac{\pi}4)$, we do not reconstruct the exchange element when the undesired parity sign is received; instead, we turn on the other two same elements. 

After applying the CNOT$^{(-)}$  {gate and restoring} the pair $(\gamma_4, \gamma_5)$, we  {obtain}
$$(|0_A1_B1_C0_D\rangle, -|0_A0_B1_C1_D\rangle,|1_A1_B1_C1_D\rangle,-|1_A0_B1_C0_D\rangle)^T.$$
 {Similarly to} before, we  {revert to} the input basis before measurement $M_2$ and  {obtain} the superposition state
\begin{equation}
-i\left(
\begin{array}{ccc}
|0_A0_B0_C0_D\rangle-|0_A1_B1_C0_D\rangle \\
|0_A0_B1_C1_D\rangle-|0_A1_B0_C1_D\rangle \\
|1_A0_B0_C1_D\rangle-|1_A1_B1_C1_D\rangle \\
|1_A0_B1_C0_D\rangle-|1_A1_B0_C0_D\rangle \\
\end{array}
\right).
\end{equation}

Finally, we  {perform} $M_2$ and restore the output fermion parity to  {that of} the input's {parity} [see Fig. \ref{fig3}(b)]. 

\subsection{Efficiency of the correction}

Since our quantum process is dissipationless and topological in  {an} ideal  {scenario}, it is fault-tolerant and very efficient when the temperature is  {low} enough. We are not going to repeat the discussions on the possible error and efficiency of the general  topological quantum computation processes \cite{TQCR,TQCR1,sparse,Bravyi}.  Efficiency here means the number and complexity of  the measurements, the signal sending, and the corrected gates. 

As mentioned before, the manipulation of moving MZMs remains a difficult task. Therefore, we focus on the case of MEMs.
 {In general, the time complexity (the number of quantum gates required) of the correction operation of the two processes we proposed , is $O(N)$, where N refers to the number of CNOT gates in the circuit. However, from the perspective of implementation, Process I is easier than  {Process} II because the structure of the phase element $G(\frac{\pi}4)$  {in Process II} is much  {more} complicated than that of  the $X$ gate  {in Process I} \cite{Zhan}.} The former involves  seven channels of chiral MEMs, and the interaction between the MEMs is specially manipulated, while the latter involves the exchange  {of} two channels of free chiral MEMs, which can be easily done by the propagation of MEMs along the interface between chiral superconductor and normal metal \cite{Zhan}.

 {Then we can evaluate the efficiency of implementing the CNOT gate. The study referenced in \cite{measure CNOT} presents a measurement-based approach for realizing the CNOT gate. According to this method, the execution of the CNOT gate involves three measurement operations alongside two quantum gates. In contrast, our proposed scheme necessitates only two measurements in addition to two quantum gate operations, with a time complexity that is comparable. However, the approach outlined in \cite{measure CNOT} necessitates the use of auxiliary qubits, which must be reinitialized following each measurement. This requirement results in a space complexity (pertaining to the computational resources consumed) of $O(N)$, where $N$ represents the quantity of CNOT gates. Our methodology, on the other hand, does not demand extra auxiliary bits, thereby maintaining a space complexity of $O(1)$.}

 {Subsequently, the efficiency of this scheme in realizing universal quantum computing can be analyzed. To achieve universal topological quantum computing, \cite{Bravyi} chose to generate a magic state, which is simulated by adding noise and then purifying the quantum state. Their analysis reveals that, beyond the topological logic gates present in the quantum circuit, supplementary non-topological computational operations are essential for emulating a quantum circuit. The non-topological computational operations necessary to simulate a quantum circuit comprising $L$ gates  {escalate to} $O(L\ln L^3)$. In our case, the non-topological operation is measurement, meaning the total number of non-topological computational operations present in our quantum circuit amounts to  {$O(N)$}, where  {$N<L$} specifically refers to the count of CNOT gates rather than the total number of logic gates in \cite{Bravyi}.
}

Besides the above discussions on the efficiency and the common error shared with any topological quantum computing process \cite{Das}, we discuss the efficiency of  the charge detection in the measurements. In the measurements $M_1$ and $M_2$, we need a charge-detecting meter to determine even or odd fermion parity. In  {the} literature, there are many proposals to design this kind of  {meter} \cite{Measure1,Measure2,Measure3,PD}. For example, Haack, F\"orster, and B\"uttiker designed a Mach-Zehnder interferometer coupled capacitively to two double quantum dots to detect the spin parity of the two qubit system \cite{PD}.  The efficiency of the parity meter is given by the rate between the inverse of the time needed by the detector to distinguish the signal from the output noise \cite{Ko,Bu} and the inverse of the coherence time of the measured quantum system.  

Such a Mach-Zehnder interferometer is widely used  in the two-dimensional electron gas in the quantum Hall regime with electron motion along edge states \cite{Hall1,Hall2,Hall3,Hall4,Hall5,Hall6}. We consider the MEMs in the chiral superconductor which are similar to the chiral edge modes in quantum Hall effects. Thus, a similar efficiency study using the Mach-Zehnder interferometer may also be done  {to detect} the fermion parity with $M_1$ and $M_2$. We will leave the detailed designation of our parity meter and the calculation of the efficiency to the further work.

\section{Conclusions}

In  {conclusion}, we have designed a set of dissipationless universal topological quantum gates with the MEMs in the chiral superconductor according to our fermion parity correction process and the probabilistic universal topological quantum gates designed in our previous work \cite{Zhan}. Our process may also be applied to the systems  {by} directly moving the MZMs. Our process is more efficient than  {the existing} parity correction process by saving the quantum information resource and reducing the number and complexity  {of measurements}.  There is also  {a} correspondence between our  process  {and} the measurement-only scheme, and our study may inspire the universal gates, devices, and  {element} designations for the measurement-only topological quantum computer. 

\acknowledgements
We acknowledge useful discussions with Xin Liu. This work  {is} supported by NNSF of China with No. 12174067.

\appendix

\section{General corrective approach}\label{app}

The corrective approach discussed above is not unique. {For example,  {without} using the SWAP gate in Fig. \ref{fig2}, after the measurement $M_1$ and the $X_C\otimes X_D$ operation, the basis becomes $(|0_A1_B0_C1_D\rangle$, $-|0_A0_B0_C0_D\rangle$, $|1_A0_B0_C1_D\rangle$, $-|1_A1_B0_C0_D\rangle)^T$.} This basis and the one in \eqref{Dbasis} only differ by a linear transformation, which can be realized using a two-qubit gate such as  {the} {$Y$ gate $Y^{(2)}=Y\oplus Y^T$, {where $Y = HZHZ$ and $Y^T=ZHZH$ (see below)}. Another realization is to first apply $X=X_A\otimes X_B$, then apply $Z=Z_B\otimes Z_C$.} 

\begin{figure}[!h]
        \centering \includegraphics[width=1\columnwidth]{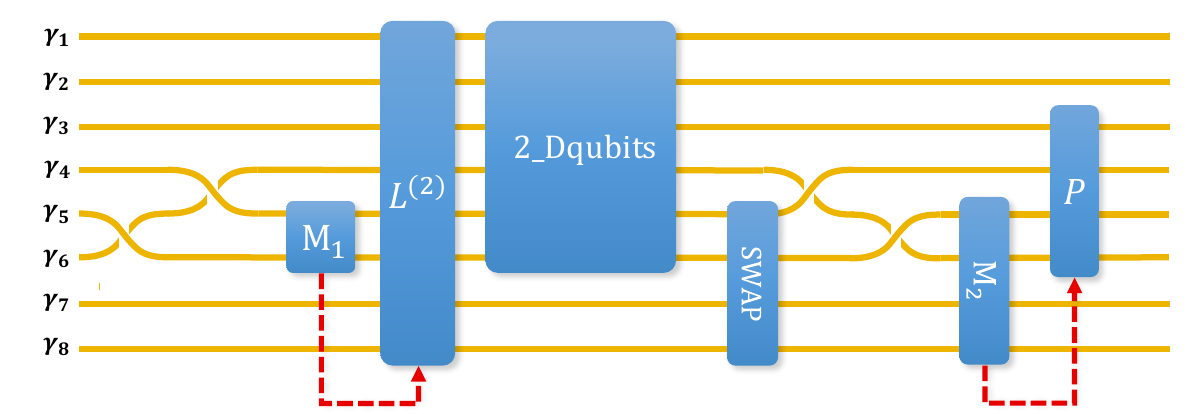}
        \caption{(Color online) Correcting the undesired fermion parity measured. $L^{(2)}$ is a possible linear transformation, and $P$ is a Pauli gate to correct the undesired fermion parity. 
        }\label{fig6}       
\end{figure}

More {generally}, for any other two-qubit quantum gate that generates entangled states (2-Dqubits gate), one can  {always} correct the undesired fermion parity using the general process depicted in {Fig. \ref{fig6}}. Here, $L^{(2)}$ represents a linear transformation used to rectify the first measurement, while $P$ denotes a Pauli gate employed to correct the undesired fermion parity observed in the second measurement.

\section{Braiding diagrams of quantum gates }\label{app}

\begin{figure}[!h]
        \centering 
        \includegraphics[width=1\columnwidth]{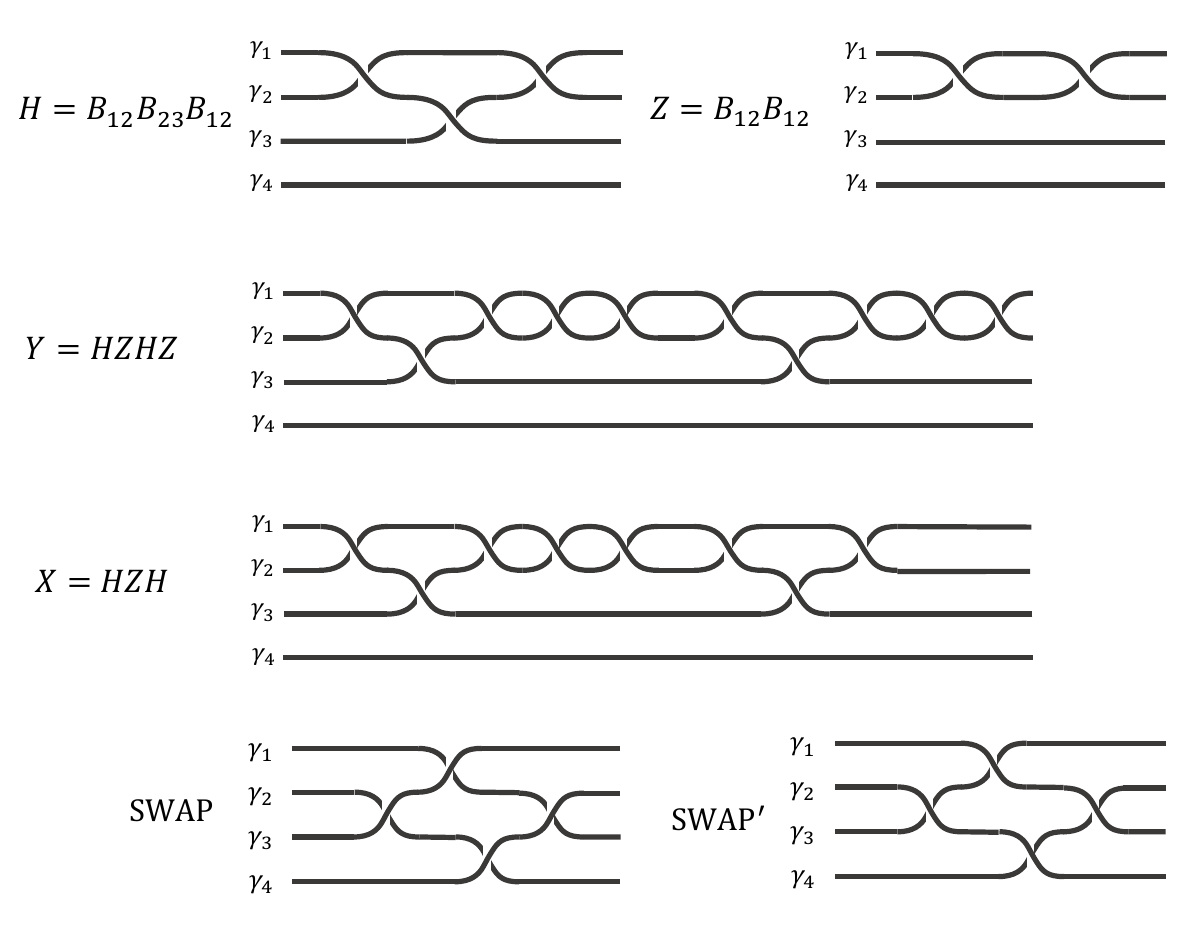}
        \caption{{Braiding diagrams} for H,Z,Y,X and SWAP gates.
        }\label{fig7}       
\end{figure}

{In this appendix, we show the braiding diagrams of the Pauli gates $X,Y,Z$ and the SWAP gate {in Fig. \ref{fig7}}. The computational basis for two SWAP gate is 
$(|0_A0_B\rangle$, $|0_A1_B\rangle$,  $|1_A0_B\rangle$, $|1_A1_B\rangle)^T$, and 
\begin{eqnarray}
 		SWAP=B_{32}B_{21}B_{43}B_{32}=
		\frac{1}{\sqrt{2}}\left(
		\begin{array}{cccc}
		1  & 0 & 0 & 0 \\
		0  & 0 & 1 & 0 \\
		0  & -1 & 0 & 0 \\
		0 & 0 & 0 & 1
		\end{array}
		\right),
		\nonumber
		\end{eqnarray}
 \begin{eqnarray}
		SWAP^{'}=B_{23}B_{12}B_{34}B_{23}=
		\frac{1}{\sqrt{2}}\left(
		\begin{array}{cccc}
		1  & 0 & 0 & 0 \\
		0  & 0 & -1 & 0 \\
		0  & 1 & 0 & 0 \\
		0 & 0 & 0 & 1
		\end{array}
		\right).
		\nonumber
		\end{eqnarray}
From the braiding diagrams, the corresponding designation of quantum gates can be realized by MEMs \cite{Zhan}.}



\section{Braiding matrix }\label{app}

In the {main text}, we use braiding operations to obtain superposition states. Here, we present the complete {matrices} of the braiding operations under even parity. {The computational basis used here is $(|0_A0_B0_C0_D\rangle$, $|0_A0_B1_C1_D\rangle$,  $|0_A1_B0_C1_D\rangle$, $|0_A1_B1_C0_D\rangle$, 
$|1_A0_B0_C1_D\rangle$, $|1_A0_B1_C0_D\rangle$, $1_A1_B0_C0_D\rangle$, $|1_A1_B1_C1_D\rangle)^T$, then the braiding operations are
\begin{eqnarray}
		(B_{45}B_{56})^{+}=
		\frac{1}{\sqrt{2}}
		\begin{pmatrix}
		i  & 0 & 0 & 1 & 0 & 0 & 0 & 0\\
		0  & -1 & i & 0 & 0 & 0 & 0 & 0\\
		0  & 1 & i & 0 & 0 & 0 & 0 & 0\\
		i & 0 & 0 & -1 & 0 & 0 & 0 & 0\\
		0  & 0 & 0 & 0 & i & 0 & 0 & 1\\
		0  & 0 & 0 & 0 & 0 & -1 & i & 0\\
		0  & 0 & 0 & 0 & 0 & 1 & i & 0\\
		0 & 0 & 0 & 0 & i & 0 & 0 & -1
		\end{pmatrix}
		,\nonumber
		\end{eqnarray}
		\begin{eqnarray}
			(B_{65}B_{54})^{+}=
		\frac{1}{\sqrt{2}}
		\begin{pmatrix}
		-1  & 0 & 0 & -1 & 0 & 0 & 0 & 0\\
		0  & i & -i & 0 & 0 & 0 & 0 & 0\\
		0  & -1 & -1 & 0 & 0 & 0 & 0 & 0\\
		-i & 0 & 0 & i & 0 & 0 & 0 & 0\\
		0  & 0 & 0 & 0 & -1 & 0 & 0 & -1\\
		0  & 0 & 0 & 0 & 0 & i & -i & 0\\
		0  & 0 & 0 & 0 & 0 & -1 & -1 & 0\\
		0 & 0 & 0 & 0 & -i & 0 & 0 & i
		\end{pmatrix}.
		\nonumber
		\end{eqnarray}} 

\begin{thebibliography}{99}

\bibitem{topo1}  {C. L. Kane and E. J. Mele, $Z_2$ Topological order and the quantum spin Hall effect, Phys. Rev. Lett. {\bf 95}, 146802 (2005).}

\bibitem{topo2}   {C. L. Kane and E. J. Mele, Quantum spin Hall effect in graphene, Phys. Rev. Lett. {\bf 95}, 226801 (2005)}.

\bibitem{topo6}  {L. Fu, C. L. Kane, and E. J. Mele, Topological insulators in three dimensions, Phys. Rev. Lett. {\bf 98}, 106803 (2007).}

\bibitem{tosp1}  M. Z. Hasan and C. L. Kane, Colloquium: Topological insulators, Rev. Mod. Phys. {\bf 82}, 3045 (2010).

\bibitem{tosp2}  X.-L. Qi and S.-C. Zhang, Topological insulators and superconductors, Rev. Mod. Phys. {\bf 83}, 1057 (2011).

\bibitem{topo3}   { C.-Z. Chang {\it et al.}, Experimental observation of the quantum anomalous Hall effect in a magnetic topological insulator, Science  {\bf 340}, 167 (2013).}



\bibitem{topo4}   {M. Gong, Y. Qian, M. Yan, V. W. Scarola, and C. Zhang, Dzyaloshinskii-Moriya interaction
and spiral order in spin-orbit
coupled optical lattices, Sci. Rep. {\bf 5}, 10050 (2015).}

\bibitem{topo5}   {M. Yan, Y. Qian, H.-Y. Hui, M. Gong, C. Zhang, and V. W. Scarola, Spin-orbit-driven transitions between Mott insulators and finite-momentum superfluids of bosons in optical lattices, Phys. Rev. A {\bf 96}, 053619 (2017).}

\bibitem{MR} G. Moore and N. Read, Nonabelions in the fractional quantum hall effect, Nucl. Phys. B {\bf 360}, 362 (1991).
\bibitem{RG2000} N. Read and D. Green, Paired states of fermions in two dimensions with breaking of parity and time-reversal symmetries and the fractional quantum Hall effect, Phys. Rev. B {\bf 61}, 10267 (2000).
\bibitem{NW} C. Nayak and F. Wilczek, 2n-quasihole states realize 2$^{n-1}$-dimensional spinor braiding statistics in paired quantum Hall states, Nucl. Phys. B {\bf 479}, 529 (1996).
\bibitem{Inv} D. A. Ivanov, Non-Abelian statistics of half-quantum vortices in $p$-Wave Superconductors, Phys. Rev. Lett. {\bf 86}, 268 (2001).
\bibitem{K1} A. Y.  Kitaev, Fault-tolerant quantum computation by anyons,  Ann. Phys. (NY) {\bf 303},   2 (2003).
\bibitem{Freedman2003} M. H. Freedman, A. Kitaev, M. J. Larsen, and Z. Wang, Topological quantum computation, Bull. Amer. Math. Soc. {\bf 40}, 31 (2003).
\bibitem{Geo} L. S. Georgiev, Topologically protected gates for quantum computation with non-Abelian anyons in the Pfaffian quantum Hall state, Phys. Rev. B {\bf  74}, 235112 (2006).
\bibitem{Geo1} L. S. Georgiev, {\it Lie Theory and Its Applications in Physics VII},  ed. V.K. Dobrev et al, Heron Press, Sofia, 2008.
\bibitem{TQCR}   C. Nayak, S. H. Simon, A. Stern, M. Freedman, and S. Das Sarma, Non-Abelian anyons and topological quantum computation, Rev. Mod. Phys.  {\bf 80}, 1083 (2008).


\bibitem{TQCR1} B. Field and T. Simula, Introduction to topological quantum computation with non-Abelian anyons, Quantum Sci. Technol. {\bf 3}, 045004 (2018).
\bibitem{K3} A. Y. Kitaev, Unpaired Majorana fermions in quantum wires, Physics-Uspekhi {\bf 44},  131(2001).

\bibitem{prox1} J.  Alicea, Y. Oreg, G. Refael, F. von Oppen and M. P.  A. Fisher, Non-Abelian statistics and topological quantum information, Nat. Phys. {\bf 7}, 412 (2011).


\bibitem{fukane} L. Fu and C. L. Kane, Superconducting proximity effect and Majorana fermions at the surface of a topological insulator, Phys. Rev. Lett. {\bf 100}, 096407 (2008).
\bibitem{Sau2010} J. D. Sau, R. M. Lutchyn, S. Tewari, and S. Das Sarma, Generic new platform for topological quantum computation using semiconductor heterostructures, Phys. Rev. Lett. {\bf 104}, 040502 (2010).
\bibitem{Das Sarma2010} R. M. Lutchyn, J. D. Sau, and S. Das Sarma, Majorana fermions and a topological phase transition in semiconductor-superconductor heterostructures, Phys. Rev. Lett. {\bf 105}, 077001 (2010).

\bibitem{Hunting} For recent development,  see review, e.g., A. Yazdani, F. von Oppen, B. I. Halperin, and A. Yacoby, Hunting for Majoranas, Science {\bf 380}, eade0850 (2023)

\bibitem{theory} Also see, Y. Tanaka, S. Tamura, and J. Cayao, Theory of Majorana zero modes in unconventional superconductors, E-print arXiv 2402.00643 (2024).

\bibitem{Das2013}  {S. Takei, B. M. Fregoso, H.-Y. Hui, A. M. Lobos, and S. Das Sarma, Soft superconducting gap in semiconductor Majorana nanowires, Phys. Rev. Lett. {\bf 110}, 186803  (2013).}

\bibitem{Sau2015}  {M. Barkeshli, and J. D. Sau, Physical architecture for a universal topological quantum computer based on a network of Majorana nanowires, arXiv:1509.07135.}


\bibitem{Mourik2012} V. Mourik, K. Zuo, S. M. Frolov, S. R. Plissard, E. P. A. M. Bakkers, and L. P. Kouwenhoven, Signatures of Majorana fermions in hybrid superconductor-semiconductor nanowire Devices, Science {\bf 336}, 1003 (2012).
\bibitem{Deng2012} M. T. Deng,  C. L. Yu, G. Y. Huang, M. Larsson, P. Caroff, and H. Q. Xu, Anomalous zero-bias conductance peak in a Nb–InSb nanowire–Nb hybrid device, Nano Lett. {\bf 12}, 6414 (2012).
\bibitem{ADas}  A. Das, Y. Ronen, Y. Most, Y. Oreg, M. Heiblum, and H. Shtrikman, Zero-bias peaks and splitting in an Al–InAs nanowire topological superconductor as a signature of Majorana fermions, Nat. Phys. {\bf 8}, 887 (2012).
\bibitem{Chur}  H. O. H. Churchill, V. Fatemi, K. Grove-Rasmussen, M. T. Deng, P. Caroff, H. Q. Xu, and C. M. Marcus, Superconductor-nanowire devices from tunneling to the multichannel regime: Zero-bias oscillations and magnetoconductance crossover, Phys. Rev. B {\bf 87}, 241401(R) (2013).
\bibitem{Deng}  M. T. Deng, C. L. Yu, G. Y. Huang, M. Larsson, P. Caroff, and H. Q. Xu, Parity independence of the zero-bias conductance peak in a nanowire based topological superconductor-quantum dot hybrid device, Sci. Rep. {\bf 4}, 7261 (2014).
\bibitem{Nadj-Perge2014} S. Nadj-Perge, Ilya K. Drozdov, J. Li, H. Chen, S. Jeon, J. Seo, A. H. MacDonald, B. A. Bernevig, and A. Yazdani, Observation of Majorana fermions in ferromagnetic atomic chains on a superconductor, Science {\bf 346}, 602 (2014).
\bibitem{Jia1} J.-P. Xu, M.-X. Wang, Z. L. Liu, J.-F. Ge, X. J.  Yang, C. H. Liu, Z. A. Xu, D. D. Guan, C. L. Gao, D. Qian, Y. Liu, Q.-H. Wang, F.-C.  Zhang, Q.-K. Xue, and J.-F. Jia, Experimental detection of a Majorana mode in the core of a magnetic vortex inside a topological insulator-superconductor Bi$_2$Te$_3$/NbSe$_2$ Heterostructure, Phys. Rev. Lett. {\bf 114}, 017001 (2015).

\bibitem{Sun2016} H. H. Sun et al., Majorana zero mode detected with spin selective Andreev reflection in the vortex of a topological superconductor, Phys. Rev. Lett. {\bf 116}, 257003 (2016).

\bibitem{Kong2019} L. Kong et al., Half-integer level shift of vortex bound states in an iron-based superconductor, Nat. Phys. {\bf 15}, 1181 (2019).
\bibitem{Chen2020} C. Chen, K. Jiang, Y. Zhang, C. Liu, Y. Liu, Z. Wang and J. Wang,  Atomic line defects and zero-energy end states in monolayer Fe(Te,Se) high-temperature superconductors, Nat. Phys. {\bf 16}, 536 (2020).

\bibitem{wang2020} Z. Wang, J. O. Rodriguez, L. Jiao, S. Howard, M. Graham, G. D. Gu, T. L. Hughes, D. K. Morr, and V. Madhavan,   Evidence for dispersing 1D Majorana channels in an iron-based superconductor, Science {\bf 367} 104 (2020).

\bibitem{Chen2019} C. Chen et al., Quantized Conductance of Majorana Zero Mode in the Vortex of the Topological Superconductor (Li$_0.84$Fe$_0.16$)OHFeSe, Chin. Phys. Lett. {\bf 36}, 057403 (2019).


\bibitem{P. Zhang2019} P. Zhang et al., Multiple topological states in iron-based superconductors, Nat. Phys. {\bf 15}, 41 (2019).
\bibitem{W. Liu2020} W. Liu et al., A new Majorana platform in an Fe-As bilayer superconductor, Nat. Comm. {\bf 11}, 5688 (2020).

\bibitem{DM1} J. D. Sau, D. J. Clarke, and S. Tewari, Controlling non-Abelian statistics of Majorana fermions in semiconductor nanowires, Phys. Rev. B {\bf 84}, 094505 (2011).

\bibitem{DM2} B. van Heck, A. R. Akhmerov, F. Hassler, M. Burrello, and C. W. J. Beenakker, Coulomb-assisted braiding of Majorana fermions in a Josephson junction array, New Journal of Physics {\bf 14}, 035019 (2012).

\bibitem{DM3} T. Hyart, B. van Heck, I. C. Fulga, M. Burrello, A. R. Akhmerov, and C. W. J. Beenakker, Flux-controlled quantum computation with Majorana fermions, Phys. Rev. B {\bf 88}, 035121 (2013).

\bibitem{DM4} X.J. Liu, C. L. M. Wong, and K. T. Law, Non-Abelian Majorana doublets in time-reversal-invariant topological superconductors, Phys. Rev. X {\bf 4}, 021018 (2014).

\bibitem{DM5} X. Liu, X. Li, D.-L. Deng, X.-J. Liu, and S. Das Sarma, Majorana spintronics, Phys. Rev. B {\bf 94}, 014511 (2016).

\bibitem{DM6}  B. J. Brown and S. Roberts, Universal fault-tolerant measurement-based quantum computation, Phys. Rev.  Research {\bf 2}, 033305 (2020).



\bibitem{DM7} M. Brooks and C. Tahan, Quantum computation by spin-parity measurements with encoded spin qubits, Phys. Rev. B {\bf 108}, 035206 (2023).

\bibitem{Rau} R. Raussendorf and H. J. Briegel,  A one-way quantum computer, Phys. Rev. Lett. {\bf 86}, 5188 (2001).

\bibitem{Rau1} R. Raussendorf, D. E. Browne, and H. J. Briegel, Measurement-based quantum computation on cluster states, Phys. Rev. A {\bf 68}, 022312 (2003).

\bibitem{Measure1} C. W. J. Beenakker, D. P. DiVincenzo, C. Emary,  and M. Kindermann, Charge detection enables free-electron quantum computation, Phys. Rev. Lett. {\bf 93}, 020501 (2004).
\bibitem{Measure2} H. A. Engel and D. Loss, Fermionic Bell-state analyzer for spin qubits, Science {\bf 309}, 586 (2005).
\bibitem{Measure3} B. Trauzettel, A. N. Jordan, C. W. J. Beenakker, and M. B\"uttiker, Parity meter for charge qubits: An efficient quantum entangler, Phys. Rev. B {\bf 73}, 235331(2006).

\bibitem{boderson} P. Bonderson, M. Freedman, and C. Nayak, Measurement-only topological quantum computation, Phys. Rev. Lett. {\bf 101}, 010501 (2008).

\bibitem{Bravyi} S. Bravyi, Universal quantum computation with the $\nu=5/2$ fractional quantum Hall state, Phys. Rev. A {\bf 73}, 042313(2006).

\bibitem{measure CNOT} O. Zilberberg, B. Braunecker, and D. Loss, Controlled-NOT gate for multiparticle qubits and topological quantum computation based on parity measurements, Phys. Rev. A {\bf 77}, 012327 (2008).

\bibitem{LL} S. Q. Zhang, J. S. Hong, Y.  Xue,  X. J.  Luo,  L. W. Yu,  X. J.  Liu, and X.  Liu, Deterministic topological quantum gates for Majorana qubits without ancillary modes, E-print: arXiv 2305.18190 (2023).


\bibitem{Measure4} C. K.  McLauchlan and B. B\'eri, Fermion-parity-based computation and its Majorana-zero-mode implementation, Phys. Rev. Lett. {\bf 128}, 180504 (2022).



\bibitem{K4} A. Y.  Kitaev, Anyons in an exactly solved model and beyond, Ann. Phys. (N.Y.) {\bf 321}, 2 (2006).

\bibitem{Luo1} X. Luo, Y. G. Chen,  Z. Wang, and Y.  Yu, Intrinsic chiral topological superconductor thin films, Phys. Rev. B {\bf 108}, 235147 (2023).

\bibitem{X.-L. Qi2010} X.-L. Qi, T. L. Hughes, and S.-C. Zhang, Chiral topological superconductor from the quantum Hall state, Phys. Rev. B {\bf 82}, 184516 (2010).
\bibitem{BL} B. Lian, X.-Q. Sun, A. Vaezi, X.-L. Qi, and S.-C. Zhang, Topological quantum computation based on chiral Majorana fermions, PNAS {\bf 115}, 10938 (2018).


\bibitem{X.-G. Wen2018} W. J. Ji, and X.-G. Wen, $\frac{1}{2}(e^2/h)$ conductance plateau without 1D chiral Majorana fermions, Phys. Rev. Lett. {\bf 120}, 107002 (2018).
\bibitem{science1} M. Kayyalha, D. Xiao, R. X.  Zhang, J. H. Shin, J. Jiang, F. Wang, Y.-F. Zhao,
R. Xiao, L. Zhang, K. M. Fijalkowski, P. Mandal, M. Winnerlein, C. Gould, Q. Li, L. W. Molenkamp, M. H. W. Chan, N. Samarth, C.-Z. Chang, Absence of evidence for chiral Majorana modes in quantum anomalous Hall-superconductor devices, Science {\bf 367}, 64 (2020).



\bibitem{Zhan} Y. M. Zhan, Y. G. Chen, B. Chen, Z. Q.  Wang, Y. Yu, and X. Luo, Universal topological quantum computation with strongly correlated Majorana edge modes, New J. Phys. {\bf 24}, 043009 (2022).

\bibitem{hukane} Y. C. Hu and C. L. Kane, Fibonacci topological superconductor, Phys. Rev. Lett. {\bf 120}, 066801 (2018).

\bibitem{sparse} S. D. Sarma, M. Freedman, and C. Nayak, Majorana zero modes and topological quantum computation, npj Quan. Inf. {\bf 1}, 15001 (2015).

\bibitem{MS} G. Moore and N. Seiberg, Classical and quantum conformal field theory, Commun. Math. Phys. {\bf 123}, 177 (1989).

\bibitem{univ} P. O. Boykin, T.  Mor, M. Pulver, V. Roychowdhury, and F.  Vatan, A new universal and fault-tolerant quantum basis, Info. Proc. Lett. {\bf 75}, 101 (2000).

\bibitem{UTQC} A. Barenco, C. H. Bennett, R. Cleve, D. P. DiVincenzo, N. Margolus, P. Shor, T. Sleator, J. A.  Smolin, and H. Weinfurter, Elementary gates for quantum computation, Phys. Rev. A {\bf 52}, 3457 (1995).

\bibitem{Das} S. Das Sarma, M. Freedman, and C. Nayak, Topologically protected qubits from a possible non-Abelian fractional quantum Hall state, Phys. Rev. Lett.
{\bf 94}, 166802 (2005).

\bibitem{PD} G. Haack, H. F\"orster, and M. B\"uttiker, Parity detection and entanglement with a Mach-Zehnder interferometer, Phys. Rev. B {\bf 82}, 155303 (2010).

\bibitem{Ko} A. N. Korotkov, Continuous quantum measurement of a double dot, Phys. Rev. B {\bf 60}, 5737 (1999).

\bibitem{Bu} M. B\"uttiker, Scattering theory of current and intensity noise correlations in conductors and wave guides, Phys. Rev. B {\bf 46}, 12485 (1992).

\bibitem{Hall1}  Y. Ji, Y. Chung, D. Sprinzak, M. Heiblum, D. Mahalu, and H.  Shtrikman, An electronic Mach–Zehnder interferometer, Nature {\bf  422}, 415 (2003).
\bibitem{Hall2}  I. Neder, M. Heiblum, Y. Levinson, D. Mahalu, and V. Umansky, Unexpected behavior in a two-Path electron interferometer, Phys. Rev. Lett. {\bf 96}, 016804 (2006).
\bibitem{Hall3}  P. Roulleau, F. Portier, P. Roche, A. Cavanna, G. Faini, U. Gennser, and D. Mailly, Direct Measurement of the Coherence Length of Edge States in the Integer Quantum Hall Regime, Phys. Rev. Lett. {\bf 100}, 126802 (2008).
\bibitem{Hall4}L. V. Litvin, A. Helzel, H.-P. Tranitz, W. Wegscheider, and C. Strunk, Edge-channel interference controlled by Landau level filling, Phys. Rev. B {\bf 78}, 075303 (2008).
\bibitem{Hall5}  E. Bieri, M. Weiss, O. G\"oktas, M. Hauser, C. Sch\"onen- berger, and S. Oberholzer, Finite-bias visibility dependence in an electronic Mach-Zehnder interferometer, Phys. Rev. B {\bf 79}, 245324 (2009).
\bibitem{Hall6} P. Roulleau, F. Portier, P. Roche, A. Cavanna, G. Faini, U. Gennser, and D. Mailly, Tuning decoherence with a voltage probe, Phys. Rev. Lett. {\bf 102}, 236802 (2009).

\end{thebibliography}
\end{document}